# Improved Ultrasound Attenuation Coefficient Estimation Using Spectral Normalization on Local Interference-Free Single-Scatterer Power Spectrum

Kun-Lin Liu
*Dept. of Electrical Engineering*
*National Tsing Hua University*
Hsinchu, Taiwan
kunlin881013@gmail.com

Yu-Heng Chen
*Dept. of Electrical Engineering*
*National Tsing Hua University*
Hsinchu, Taiwan

Chiao-Yin Wang
*Dept. of Medical Imaging and*
*Radiological Sciences, College of*
*Medicine, Chang Gung University*
Taoyuan, Taiwan

Po-Hsiang Tsui
*Dept. of Medical Imaging and*
*Radiological Sciences*
*Chang Gung University*
Taoyuan, Taiwan
*Department of Medical Imaging and*
*Intervention*
*Chang Gung Memorial Hospital at*
*Linkou*
Taoyuan, Taiwan
*Institute for Radiological Research,*
*Chang Gung University and Chang*
*Gung Memorial Hospital at Linkou*
Taoyuan, Taiwan

Meng-Lin Li*
*Dept. of Electrical Engineering*
*National Tsing Hua University*
Hsinchu, Taiwan
*Inst. of Photonics Technologies*
*National Tsing Hua University*
Hsinchu, Taiwan
*Brain Research Center*
*National Tsing Hua University*
Hsinchu, Taiwan
mlli@ee.nthu.edu.tw

*Abstract*—Ultrasound attenuation coefficient estimation (ACE) can be utilized to quantify liver fat content, offering significant diagnostic potential in addressing the growing global public health issue of non-alcoholic fatty liver and other chronic liver diseases. Among ACE methods, the reference frequency method (RFM) proposed recently possesses the advantages of being system-independent and not requiring reference phantom. However, the presence of large oscillations in frequency power ratio decay curves leads to unstable ACE results with RFM, originating from noise as well as constructive and destructive interference in the backscattered signals' power spectrum. To address this issue, we propose an improved RFM version where a single-scatterer power spectrum estimator is incorporated to restore interference free single-scatterer power spectrum, thereby reducing oscillations in the frequency power ratio decay curves and greatly improving the accuracy of ACE.

## I. INTRODUCTION

Non-alcoholic fatty liver disease (NAFLD) is globally the most prevalent liver disorder. The formation of NAFLD is primarily associated with the accumulation of triglycerides. The liver, a multifunctional organ in the human body, plays a crucial role in fat metabolism. As fats from food break down into fatty acids, these acids are transported to the liver for metabolism and subsequently transformed into cholesterol and triglycerides. However, when the intake of fats exceeds the metabolic capacity of liver cells, these excess fats accumulate outside the liver cells, leading to the development of fatty liver. This condition is very common due to its widespread impact on people's health. NAFLD is now a central topic in liver research.

Among NAFLD patients, approximately 20-30% [1] will progress to more severe hepatitis, which can lead to liver fibrosis, cirrhosis, liver failure, or hepatocellular carcinoma. As the disease advances to this stage, both the patient's quality of life and life expectancy are significantly affected. Given this, early detection of NAFLD is critical, as early diagnosis and treatment can effectively prevent subsequent complications.

In modern medical practice, liver biopsy [2] is considered the "gold standard" for the diagnosing and monitoring of liver diseases. This procedure involves using a needle to extract a sample of liver tissue, which is then analyzed under a microscope to determine the cause of the disease and the extent of fibrosis. During this process, ultrasound or CT scans are typically used for guidance. However, due to its invasive nature, it's not suitable for frequent examinations. On the other hand, MRI's PDFF [3] can evaluate the fat content within the liver, providing shorter imaging times and lower costs. Nevertheless, MRI machines are substantial in size and require specialized medical personnel for operation. Thus, in some remote areas, the lack of such equipment and expertise limits the clinical use of MRI.

Therefore, there is an urgent need for a low-cost, highly accurate, and widely applicable diagnostic method in clinical practice. Previous research has demonstrated a strong positive correlation between the fat content in the liver and ultrasound attenuation. Currently, a solution involves the use of a machine called Fibroscan [4]. It evaluates liver stiffness and fat content by measuring the propagation speed of ultrasound waves within the liver. However, a significant drawback of Fibroscan is the absence of B-mode imaging support during the examination. Our objective is to utilize B-mode imaging to calculate attenuation. Fundamentally, ultrasound is non-invasive and poses no adverse effects on the human body, making it an ideal diagnostic tool.

Among various ACE methods, the Reference Frequency Method (RFM) [5] is widely regarded as the suitable method for estimating attenuation coefficients, with the advantage of being system-independent. However, RFM still has certain some challenges. For example, consider Fig. 1. The large oscillations in the frequency power ratio decay curves may render ACE unstable. This issue arises because RFM assumes that there is only a single scatterer within the sample volume at each depth. However, in reality, there are multiple scatterers within the sample volume, resulting in multi-scatterer interference in the local power spectrum. Such interference leads to large oscillations in the frequency power ratio decay curves and thus unstable ACE.

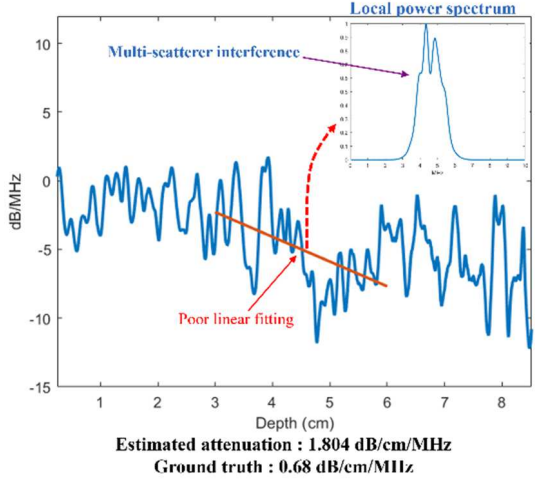

Fig. 1. Problems faced by RFM

To address this issue, our study proposes an improved version of RFM, aiming to suppress interference in power spectrum estimation, reduce oscillations, and thus obtain more accurate attenuation coefficients.

## II. MATERIALS AND METHODS

### A. Reference Frequency Method [5]

In ultrasound imaging, the power spectrum of the backscattered radiofrequency (RF) signals $S(f_i, z_k)$ can be modeled as follows (see Eq. (1)).

$$S(f_i, z_k) = G(f_i) \cdot TGC(z_k) \cdot D(f_i, z_k) \cdot BSC(f_i) \cdot A(f_i, z_k) \quad (1)$$

where $G(f_i)$ is the transmit and receive transducer responses. $TGC(z_k)$ is time gain compensation. $D(f_i, z_k)$ is the effects of focusing, beamforming, and diffraction. $BSC(f_i)$ is the backscatter coefficient which is assumed to be uniform in the local region of interest (ROI). $A(f_i, z_k)$ is the frequency-dependent attenuation, as shown in (2).

$$A(f_i, z_k) = \exp(-4\alpha f_i z_k) \quad (2)$$

RFM achieves system independence by dividing adjacent ultrasonic signal frequency components to eliminate system effects, such as focusing and time gain compensation, as shown in (3). [5] assumes that the differences of beamforming and diffraction effects between $f_i$ and $f_{i-1}$ are negligible, i.e., $D(f_i, z_k) = D(f_{i-1}, z_k)$. In addition, $TGC(z_k)$ is assumed to be independent of $f_i$. Finally, both the numerator and the denominator's $D$ and $TGC$ can be cancelled out.

$$R_s(f_i, z_k) = \frac{S(f_i, z_k)}{S(f_{i-1}, z_k)}$$
$$= \frac{G(f_i) \cdot BSC(f_i) \cdot A(f_i, z_k)}{G(f_{i-1}) \cdot BSC(f_{i-1}) \cdot A(f_{i-1}, z_k)} \quad (3)$$

After taking the natural logarithm of both sides in Eq. (3), we can obtain the linear relationship between the frequency power ratio $ln[R_s(f_i, z_k)]$ and imaging depth $z_k$, as shown in (4).

$$ln[R_s(f_i, z_k)] = \left[ln\frac{G(f_i)}{G(f_{i-1})} + ln\frac{BSC(f_i)}{BSC(f_{i-1})}\right] - 4\alpha(f_i - f_{i-1})z_k \quad (4)$$

Then the attenuation coefficient can be estimated from the slope of the decay trend of frequency power ratio, as shown in (5).

$$\alpha_i = -\frac{slope}{4(f_i - f_{i-1})z} \quad (5)$$

### B. Reconstruction of Local Interference Free Single Scatterer Power Spectrum

However, RFM assumes a single scatterer in space, whereas in reality, multiple scatterers exist within the sample volume. The received signal can be modeled as the convolution of the ultrasonic pulse with the scatterers. In the frequency domain, this corresponds to a multiplication operation, making the originally smooth power spectrum rough and causing excessive oscillations in frequency power ratio decay curves. To tackle this challenge, our study uses homomorphic filtering [6] to reconstruct a local interference free single scatterer power spectrum, which utilizes a simple deconvolution algorithm to estimate accurate spectrum, thereby reducing oscillations in the frequency power ratio decay curves and improving the accuracy of ACE. The detailed algorithm steps can be seen in Fig. 2 and are described below.

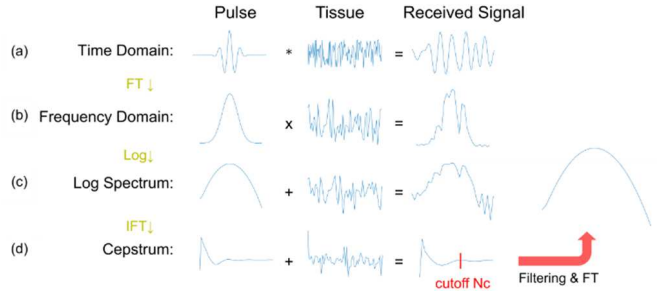

Fig. 2. Steps of the algorithm

We can model the received signal as shown in (6), which corresponds to Fig. 2(a), where $x(t)$ represents the pulse emitted by the transducer, $e(t)$ is the characteristic of the adjacent tissues that may introduce noise (e.g. reverberation and scattering from other tissues), and $y(t)$ is the back-scattered signal received by the transducer, influenced by noise. This signal is a convolution of the ultrasound pulse with tissue scatterers in the time domain.

$$y(t) = x(t) * e(t) \quad (6)$$

In the frequency domain, the property of convolution becomes multiplication, as shown in (7). This corresponds to Fig. 2(b). We can observe that the influence of reverberation and scattering from other tissues causes a fluctuated spectrum.

$$Y(f) = X(f) \times E(f) \quad (7)$$

To transform the multiplicative character of the tissue into additive, we take the logarithm, as shown in (8). Then perform inverse Fourier transform to obtain the cepstrum, as shown in (9). This corresponds to Fig. 2(c) and 2(d).

$$\log|Y(f)| = \log|X(f)| + \log|E(f)| \quad (8)$$

$$C_y = F^{-1}\{\log|Y(f)|\} \quad (9)$$

We can observe that the cepstrum of the pulse contains most of its energy in the initial samples, while the cepstrum of the signal of the scatterers is evenly distributed across cepstrum range, as anticipated. Then we select a suitable cutoff $N_c$ [6] in $C_y$ for filtering to obtain $C_p$, aiming to eliminate multiple scatterer interference so that an interference free single scatter power spectrum can be reconstructed, as shown in (10).

$$Y_p(f) = F\{C_p\} \quad (10)$$

Finally, through the filtered $Y_p$, we can obtain a smooth power spectrum. By replacing the $S$ in RFM with $Y_p$, we can further enhance the accuracy of ACE.

## III. SIMULATION RESULT

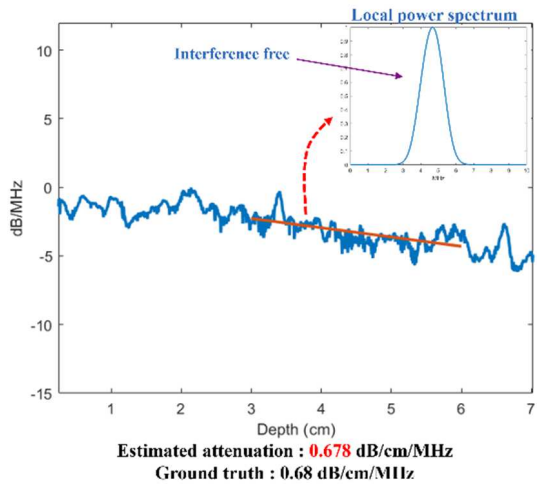

Fig. 3. Results of our proposed method

After applying our proposed processing, it is evident from Fig. 3 that the fluctuations in the frequency power ratio have significantly decreased. This reduction is attributed to our method yielding a smooth spectrum and leading to a more accurate estimation of the attenuation coefficient. Compared to Fig. 2, we can also obtain a more accurate attenuation coefficient. In the simulation, the ground truth for the attenuation coefficient is 0.68 dB/cm/MHz. If using the traditional RFM, the larger oscillation can lead to a wrong estimation of the attenuation coefficient, which is 1.804 dB/cm/MHz. However, by using our improved RFM, not only do the oscillations in the frequency power ratio decay curves decrease but also the estimated value is more accurate, at 0.678 dB/cm/MHz.

## IV. DISCUSSION AND CONCLUSIONS

Ultrasound attenuation coefficient estimation (ACE) is crucial for diagnosing non-alcoholic fatty liver disease (NAFLD) and related liver conditions. The Reference Frequency Method (RFM), while system-independent, suffers from oscillations in frequency power ratio decay curves due to interference in the power spectrum.

Our enhanced RFM addresses this by incorporating a single-scatterer power spectrum estimator, leading to the restoration of an interference-free power spectrum and reduction of oscillations. Via the application of homomorphic filtering, we achieved a more precise power spectrum.

Simulation results underscore the effectiveness of our approach. Compared to traditional RFM, our refined approach not only minimizes frequency power ratio fluctuations but also yields attenuation coefficient estimations closer to the simulation's ground truth. In essence, our improved RFM promises better accuracy and reliability in clinical applications for liver condition diagnosis.

## ACKNOWLEDGMENTS

This work is supported by National Science and Technology Council, Taiwan under the grant number NSTC 110-2221-E-007-011-MY3.